# SiPM-based neutron Anger camera with auto-calibration capabilities

A. Morozov[a,1], J. Marcos[a], L. Margato[a], D. Roulier[b] and V. Solovov[a]

[a] LIP-Coimbra, Departamento de Física, Universidade de Coimbra
Rua Larga, 3004-516 Coimbra, Portugal

[b] ILL-Institut Laue-Langevin
71 Avenue des Martyrs - CS 20156 - FR-38042 GRENOBLE CEDEX 9, France

[1] Email: andrei@coimbra.lip.pt

## Abstract

We present characterization results of a neutron Anger camera based on a lithium-6 loaded cerium activated silicate glass scintillator (33.3 x 33.3 x 1 $mm^3$) and an array of 64 silicon photomultipliers. Reconstruction of the scintillation events is performed with a statistical method, implemented on a graphics processing unit (GPU). We demonstrate that the light response model of the detector can be obtained from flood irradiation calibration data using an unsupervised iterative procedure. The useful field of view is 28 x 28 $mm^2$. The spatial resolution measured at 2.5 Å neutron beam is better than 0.6 mm FWHM and the energy resolution at the neutron peak is 11%.

## 1 Introduction

The first position sensitive neutron detector based on the Anger camera [1] principle was introduced in 1981 [2]. The detector featured a 220 mm diameter lithium-6 loaded glass scintillator and the light was read out with an array of 19 photomultiplier tubes (PMT). The positions of scintillation events were reconstructed using the center of gravity (CoG) algorithm and the reported spatial resolution was between 2 and 3 mm. Modern neutron Anger cameras with a large sensitive area are able to provide significantly better resolution. For example, a value of 1.2 mm was reported for a detector with a sensitive area of 157 x 157 $mm^2$ [3]. For small field of view cameras, sub-millimeter resolution was demonstrated (see, e.g., [4]). In part, the progress was achieved due to the use of flat panel multianode PMT assemblies instead of individual PMTs, allowing to pack more sensors per unit area and to reduce the dead space between them.

After silicon photomultipliers (SiPMs) become available, this type of light sensors was also considered for neutron Anger cameras readout [5-7]. SiPMs offer several advantages over PMTs, such as insensitivity to magnetic fields, low operating voltages (~25 V instead of ~1000 V), compactness and robustness. However, application of SiPM in neutron detectors also raises a question on their radiation hardness. Several studies investigating this aspect were performed (see, e.g., [5, 8-11]), involving both cold and fast neutrons. The results suggest that for applications in small angle neutron scattering (SANS) experiments, a SiPM lifetime of up to 10 years can be expected [5].

Reconstruction of the positions of a scintillation event in Anger camera-type detectors is most commonly performed with the CoG method. However, this method introduces systematic distortions quite severe at the camera periphery and results in the spatial resolution worse than the theoretical limit defined by the signal statistics [12]. Another approach is to apply statistical reconstruction methods [12, 13], which have several advantages, such as potentially distortion-free position reconstruction, better spatial resolution and a capability to discriminate between single- and multi-vertex scintillation events of the same total energy. The main challenges to apply the statistical methods are high computation costs and a requirement to have a mathematical model of the light response for each light sensor. This model is usually parameterized using a set of so called light response functions (LRF), each describing the response of a sensor as a function of the position of an isotropic point light source.

The LRFs can be obtained from Monte Carlo simulations, however, in this case their accuracy strongly depend on the assumptions of the simulation model and the knowledge of the optical properties for the materials of the detector, which is difficult to obtain with the required precision. Alternatively, the LRFs can be computed from calibration data acquired by scanning the field of view of the detector with a pencil beam (see, e.g., [14]). However, such calibrations are time consuming, require direct access to the detector and have to be performed on a regular basis.

In this paper we describe an early prototype of a neutron Anger camera based on an array of 64 SiPMs and a lithium-6 loaded cerium activated silicate glass scintillator. Statistical position reconstruction is performed with the maximum likelihood (ML) method. It is implemented on a graphics processing unit (GPU), resulting in the processing time per event on the order of 1 μs on a modern consumer-grade personal computer. The LRFs are obtained experimentally from flood field irradiation data by using an iterative procedure [15-18], avoiding the need for scan-based calibration. The entire cycle of the detector calibration can be performed in unsupervised mode giving the detector auto-calibration capability. We also present results of the prototype characterization performed at 2.5 Å neutron beam at ILL, including the useful field of view and the resolution (both spatial and energy) of the prototype.

## 2 Methods

### 2.1 Detector design

The neutron sensitive element of the camera prototype is a 1 mm thick, 33.3 × 33.3 mm$^2$ lithium-6 loaded cerium activated glass scintillator (GS20 from Scintacor). According to the manufacturer, this thickness results in 73% interaction probability for 2.5 Å neutrons. Capture of a neutron by a lithium-6 nucleus results in the nuclear reaction

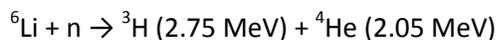
$^6$Li + n → $^3$H (2.75 MeV) + $^4$He (2.05 MeV)

The reaction products deposit their energy within short (<150 μm [7]) distance from the capture position resulting in emission of about 6200 photons with the emission peak at 395 nm.

On one side, the scintillator is coupled to a 2 mm thick soda lime glass lightguide of the same size. The side walls of the crystal and the lightguide are covered first by a layer of PTFE tape and then by a layer of aluminium foil. On the other side, the scintillator is facing an aluminium reflector. The reflector's surface was polished and then treated with a fine-thread metallic sponge to maximize diffuse light scattering. A silicone optical grease (BC-630 from Saint-Gobain) is used to couple all optical components. The refractive index of the grease is 1.47, which is very close to the refractive indices of the scintillator (1.55) and the lightguide (1.52).

The camera is equipped with 64 SiPMs by using four units of SensL ArrayC-30035-16P-PCBs. Each unit holds 4 × 4 30035 C-series SiPMs sensors with 3 x 3 $mm^2$ sensitive area, packed with a 4.2 mm pitch on a PCB holder. The frameless design allows to position arrays in such a way that the distance between all neighboring SiPMs of the camera is the same (see figure 1). The SiPMs are operated at 2.5 V above the breakdown voltage. For this value of overvoltage the manufacturer provides a photon detection efficiency of 41% at the sensitivity peak of 425 nm. The effective light detection efficiency, taking into account the GS20 emission spectrum, is about 25%.

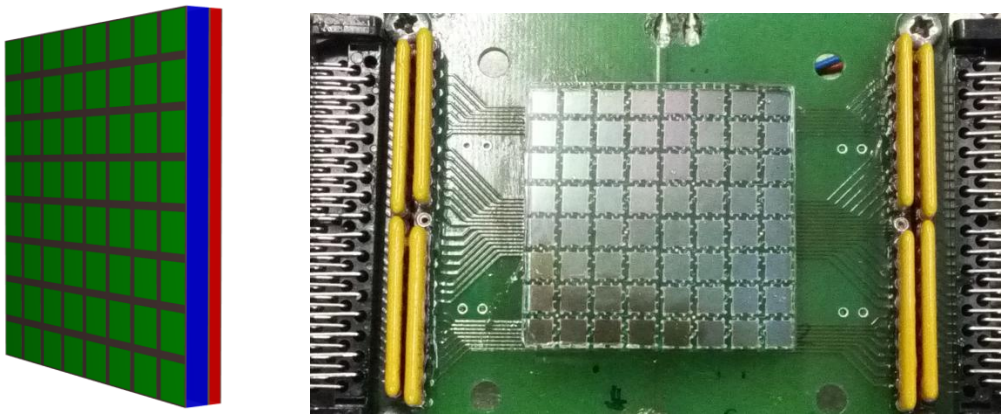

Figure 1. Left: Schematic drawing of the detector. 64 SiPMs are interfaced through a glass lightguide (2 mm thick) to the GS20 scintillator (1 mm thick). The back surface of the scintillator and the lateral surfaces of both the lightguide and the scintillator are covered with a light reflecting layer. Right: Photograph of the detector prototype before installation of the light reflective elements.

## 2.2 Readout system

The SiPMs are connected to low-noise trans-impedance amplifiers, which limit the signal bandwidth to reduce noise and prevent aliasing. The output of the amplifiers are fed into a 40-MHz 10-bit waveform digitizer. The data acquisition system is based on the TRB3 board developed by the GSI institute [19]. The waveform digitizers are implemented as two (32 channels each) add-ons for the TRB3 board. The analog sum signal from all 64 channels is fed to one of the ADC inputs. Trigger is generated internally by the TRB3 board when the sum signal surpasses a given (adjustable) level. Waveforms of 1 µs of total duration are recorded and sent over a gigabit Ethernet connection to a personal computer, where they are integrated to compute the total charge generated by each individual SiPM.

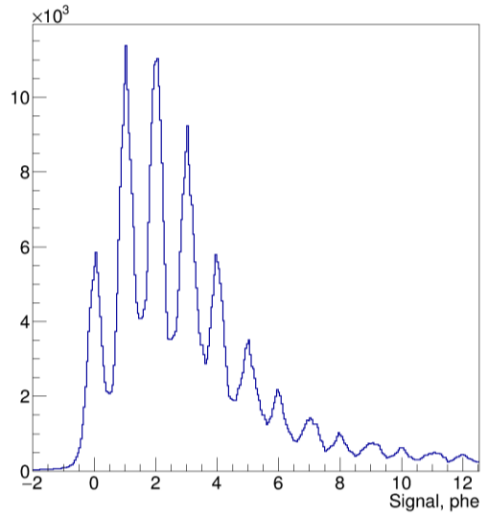

Figure 2. Distribution of the charge recorded for one of the SiPMs. The charge is converted to the number of photo-electrons.

The charge distributions recorded for all SiPMs clearly show contributions from individual photo-electrons (see figure 2). Well-defined position of the first peak and the separation between the peaks provide the data needed to convert measured charge to the corresponding number of photo-electrons.

The mean signal due to the SiPM dark counts during the 1 μs measurement window is ≈0.25 photo-electrons. This value, measured without beam and using triggering from an external pulse generator, is consistent with the 300 kHz dark count rate given by the manufacturer.

## 2.3 Experimental setup and procedures

The camera prototype was mounted inside a light-tight enclosure with a 1 mm thick aluminium entrance window and installed at the CT1 beam line of ILL (2.5 Å neutron wavelength) at normal incidence to the beam.

The entire sensitive area of the camera was irradiated to record flood field datasets. The beam had a noticeable degree of non-uniformity, which results in reduced intensity in the top-right corner of the reconstructed images.

Several datasets were recorded with 1 mm thick cadmium masks. The first mask, used to assess reconstruction fidelity, has an array of 0.4 mm diameter holes drilled with a pitch of 2 mm in both X and Y directions. The second one, used to measure the spatial resolution, has a 0.4 wide and 40 mm long slit machined in it. All measurements were performed with the masks installed inside the camera enclosure in direct contact with the light reflector.

## 2.4 Scintillation event reconstruction

Reconstruction of the position and energy (defined here as the number of emitted photons) of a scintillation event is performed by the maximum likelihood (ML) method assuming Poisson distribution of the number of photo-electrons in each channel. The LRFs, parameterized with B-splines [20], are computed using an iterative method (see the next section) from a flood field irradiation dataset. The search for the best match between the experimental and the predicted distributions of the number of photo-electrons across the SiPM array is performed using the contracting grids algorithm implemented on GPU. The initial grid is centered at the position found by the CoG method. A linear "stretch" (factor of 1.5) from the camera center is applied to the CoG reconstructed positions to compensate for the strong distortion introduced by that method. A detailed description of the reconstruction procedure can be found in [18].

## 2.5 Iterative reconstruction of LRFs

The iterative LRF reconstruction method has already been described in detail in our previous publications [15-18]. The method requires two datasets for calibration events, distributed over the field of view of the detector: one with the SiPM signals and the other one with the estimates of the event positions. The iterative cycle consists of two stages: during the first stage the signals and the event positions are used to evaluate the LRFs of the SiPMs. In the second stage, the new estimated event positions are obtained with the ML method using these LRFs. The cycle is repeated until convergence is reached in a parameter proportional to the chi-square of event position reconstruction averaged over all events [18].

The iterative procedure starts from the LRF reconstruction stage using position estimates given by the CoG reconstruction. During the first three iterations, an LRF parameterization scheme is used in which all sensors share the same LRF profile, but have individual scaling factors to account for the variations in the detection efficiency of the SiPMs. This technique allows to establish, during these first iterations, the general profile of the LRFs. For the next iterations the parameterization scheme is switched to the one in which each sensor has an individual LRF in order to take into account the differences between the individual sensors, e.g., due to the difference in the contribution from scattered light.

Typically convergence is reached after 10 iterations. The extraction of the signal per photo-electron value for each SiPM channel and the iterative LRF reconstruction procedure are implemented in scripts. The entire cycle of the camera calibration can be performed fully automatically using a flood field irradiation dataset. For $5 \cdot 10^5$ events the iteration procedure takes less than a minute on a consumer-grade personal computer equipped with an NVIDIA GTX 1070 GPU. All the tools required for the calibration and the reconstruction procedures are implemented in an open source toolkit ANTS2 [21, 22].

## 3 Results and discussion

Figure 3 shows the results of position reconstruction of a flood irradiation dataset performed with the CoG (left) and the ML (right) methods. The LRFs are obtained in the iterative procedure described above.

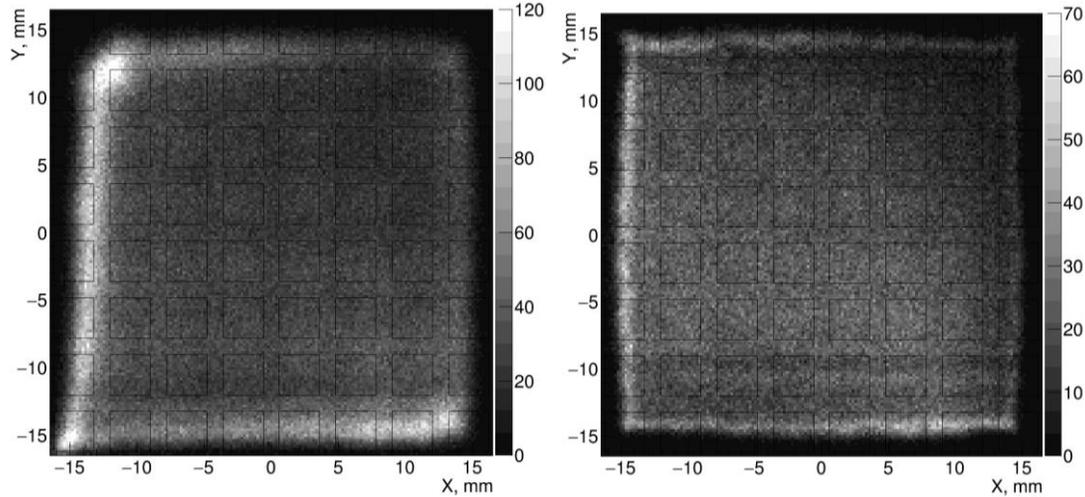

Figure 3. Reconstruction of a flood irradiation dataset performed with the CoG (left) and ML (right) methods. The image area matches the scintillator dimensions. The contours of the SiPM sensitive areas are shown by thin black lines.

The results of position reconstruction for a dataset recorded with the cadmium mask with a grid of holes are shown in figure 4 (CoG and ML) and the corresponding projections along the X and Y axes for the ML reconstruction are presented in figure 5. As one can see from figures 4 (right) and 5, the ML reconstruction in the central area of 28 x 28 mm$^2$ is quite precise. In the area of 2 mm from the periphery, the reconstructed events either fail the chi-square discrimination [18] or appear wrongly reconstructed in a 1 mm-wide area at ≈15 mm from the center, hence the density of events is higher there (see figure 3, right). This reconstruction pattern is similar to what was observed for the compact gamma camera in our previous study [18].

The ML reconstruction at the border fails since we assume axial symmetry of the LRFs. While this assumption is necessary to apply the iterative LRF reconstruction procedure [18], the LRFs at the periphery are not axially-symmetric due to strong light scattering on the lateral surfaces of the scintillator and the lightguide. In the present study the distortions are even stronger since the lateral surfaces are covered with a light-reflecting layer in contrast to the black walls of the gamma camera [18].

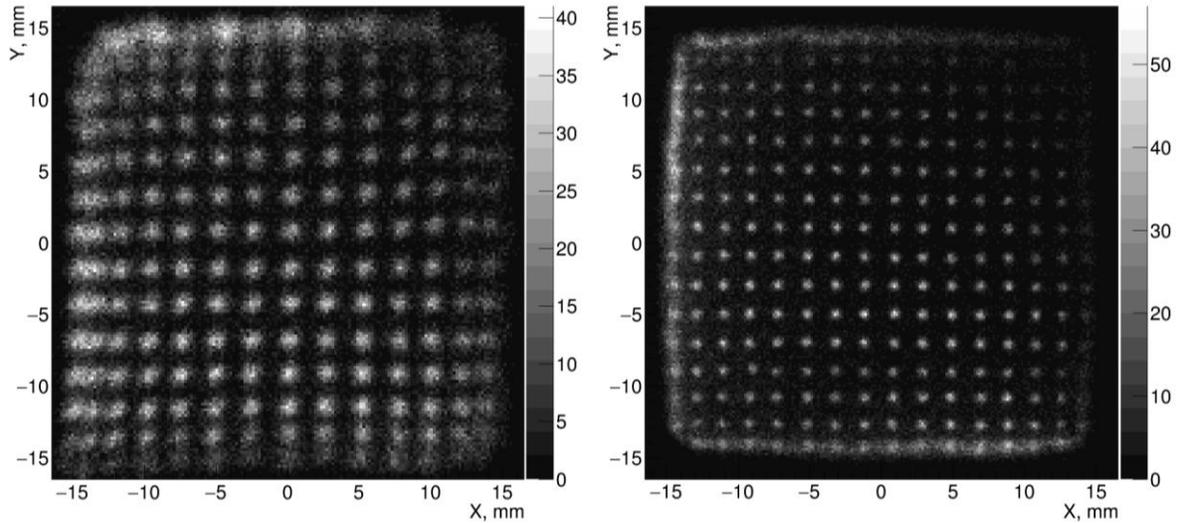

Figure 4. CoG (left) and ML reconstruction of a dataset recorded with the cadmium mask with a grid of holes (0.4 mm diameter and 2 mm pitch). The image area matches the scintillator dimensions.

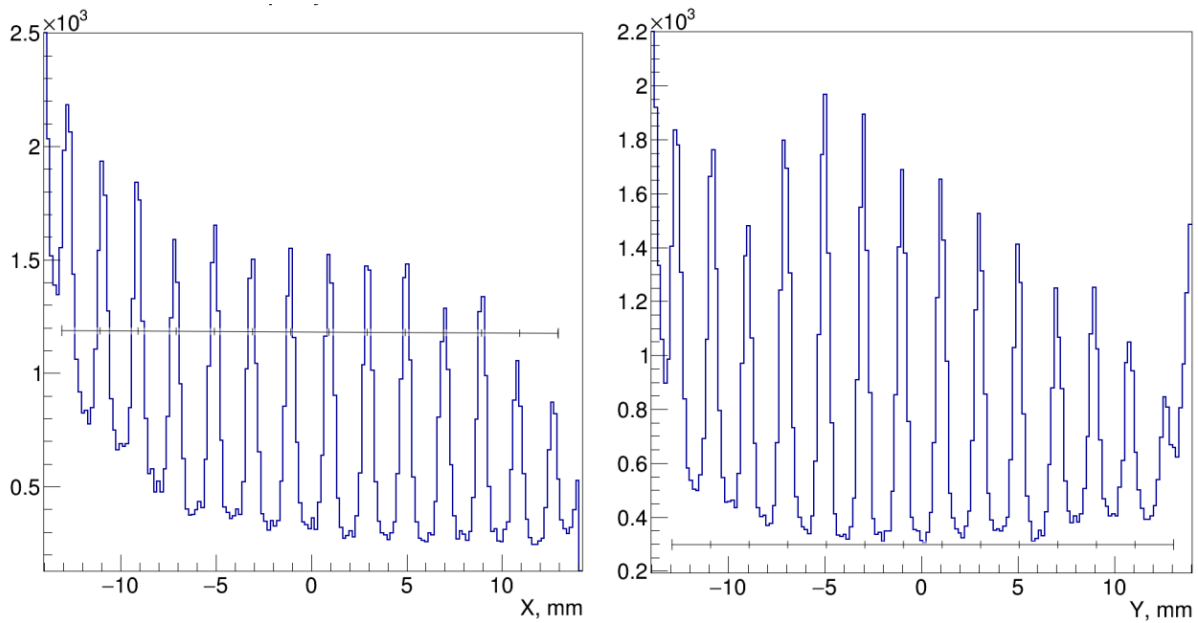

Figure 5. Y-projection along X (left) and the X-projection along Y (right) of the central 28 x 28 mm$^2$ area of the ML reconstructed image shown in figure 4 (right). The tick marks of the thin line have a pitch of 2 mm.

Despite the presence of the reflective layer at the lateral surfaces of the scintillator and the lightguide, all SiPMs except the ones situated at the border exhibit axially-symmetric response. An example of the average signal versus reconstructed position (flood field dataset) for a SiPM in the second row from the top is shown in figure 6 (left). Figure 6 (right) demonstrates that the reconstructed LRF of this SiPM describes the experimental data quite well. Note that the LRF value far from the sensor center shows a level of approximately 2 photo-electrons, which is larger than the background due to the dark counts (≈0.25 photo-electrons).

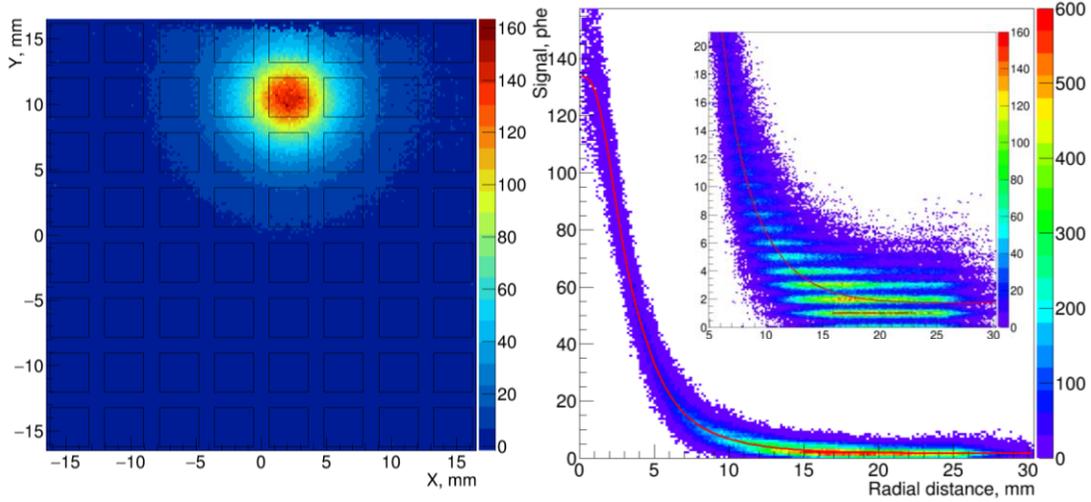

Figure 6. Left: Signal (in photo-electrons) of a SiPM (5th to the right, 2nd row down from the top-left corner) as a function of the reconstructed event position. The contours of the SiPM sensitive areas are shown by the thin black lines. Right: Signal of the same SiPM (color coded) versus the distance between the sensor center and the reconstructed event position. The iteratively-reconstructed LRF curve is shown by the red line. The insert shows the signal values below 20 photo-electrons.

The reflective layer on the lateral surfaces is introduced to improve uniformity of the sum signal of all SiPMs (and the reconstructed energy) over the entire field of view of the detector, targeting good energy resolution. For the flood field dataset, the mean sum signal (64 SiPMs) of the neutron peak is 650 photo-electrons, which includes ≈16 photo-electrons due to the dark counts. The distribution of the sum signals and the ML-reconstructed event energy is shown in Figure 7.

In order to calculate the energy resolution from the sum signal distribution, the contribution of the dark counts have to be taken into account. Subtracting 16 photo-electrons from the peak position, the energy resolution obtained for the entire sensitive area and the central 28 x 28 mm$^2$ area is 12.2% and 11.7%, respectively. The energy resolution values given by the ML method are slightly better (11.5% and 10.7%), and the peak is more symmetric. We attribute this improvement to the fact that the iteratively reconstructed response takes into account differences in the photon detection efficiency and the crosstalk of individual SiPMs. The obtained energy resolution is better than the value given by the GS20 manufacturer (14%), but matches the energy resolution (11%) recently reported for a study with a GS20 scintillator and PMT readout [23].

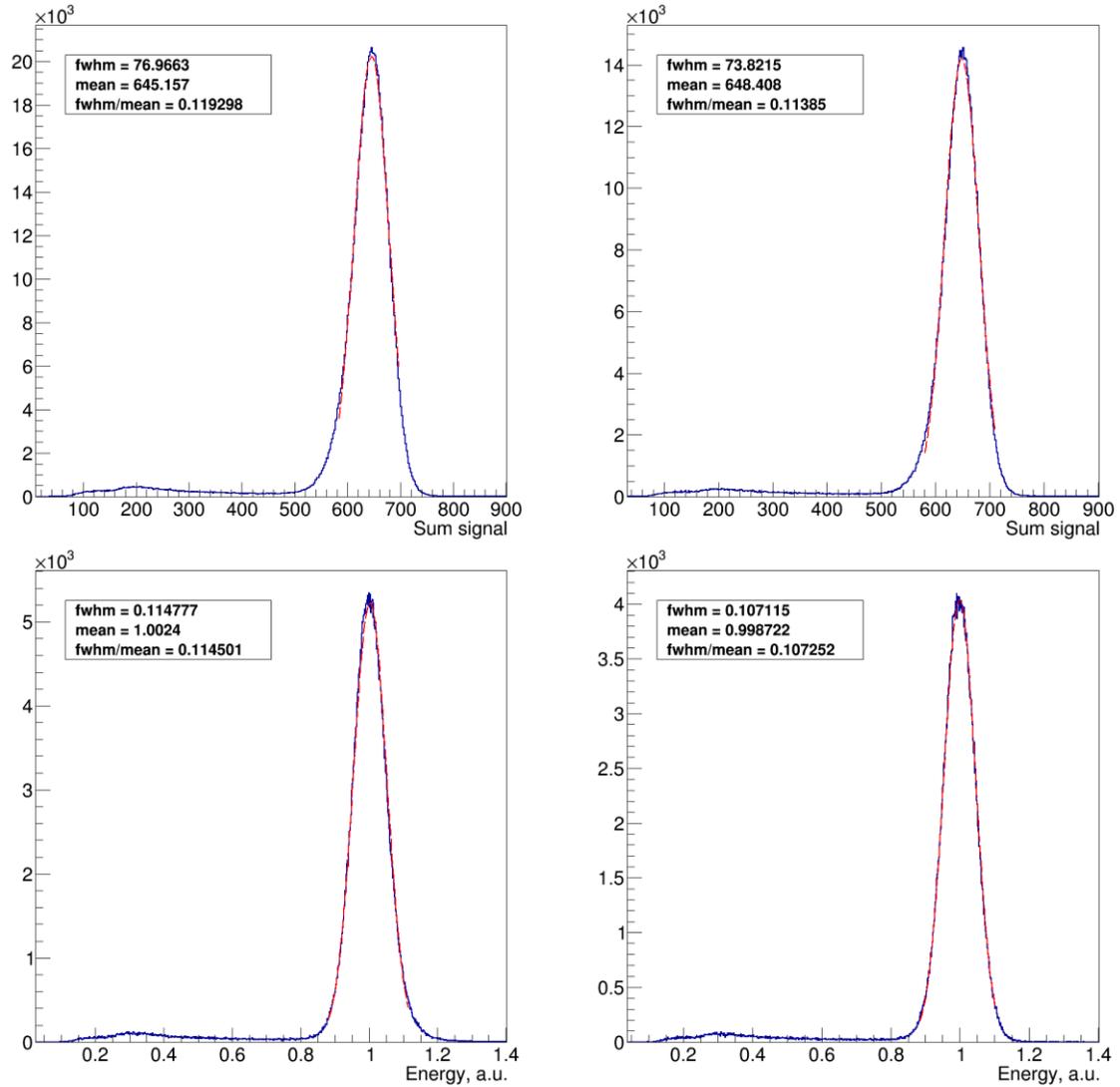

Figure 7. Distribution of the sum signal (top row) and the event energy reconstructed with the ML method (bottom row) for the entire flood field dataset (left column) and the events reconstructed in the central area of 28 x 28 mm$^2$. The Gaussian fit of the neutron peak is shown by the dashed line. The text boxes show the corresponding peak position and the FWHM value given by the fit.

Figures 8 and 9 show ML reconstructed images for datasets recorded with a cadmium mask with a 0.4 mm wide slit, and the corresponding projection profiles in the direction orthogonal to the slit. The Gaussian fits of the profiles give FWHM of ≤ 0.60 mm. Deconvolution of the slit width suggests that the intrinsic resolution of the detector is better than 0.53 mm.

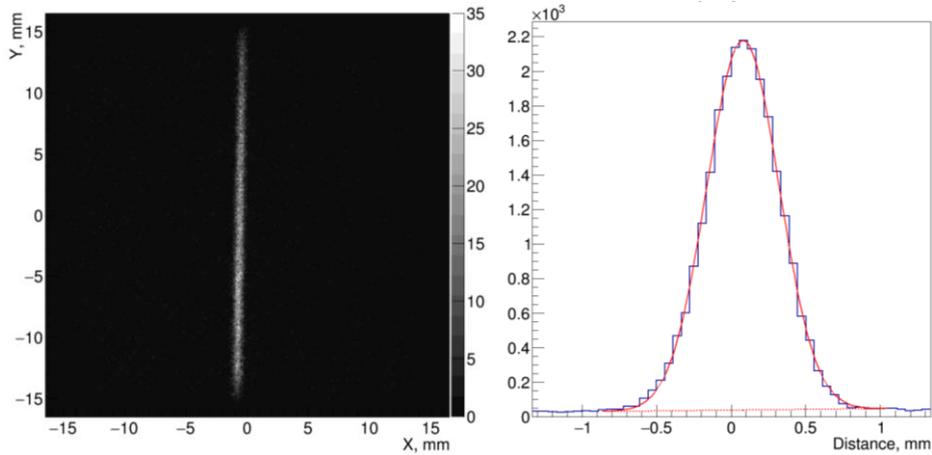

Figure 8. Left: Reconstructed image (ML method) for the dataset recorded with the cadmium mask with a 0.4 mm wide slit. Right: The Gaussian fit of the image projection in the direction orthogonal to the slit gives 0.57 mm FWHM.

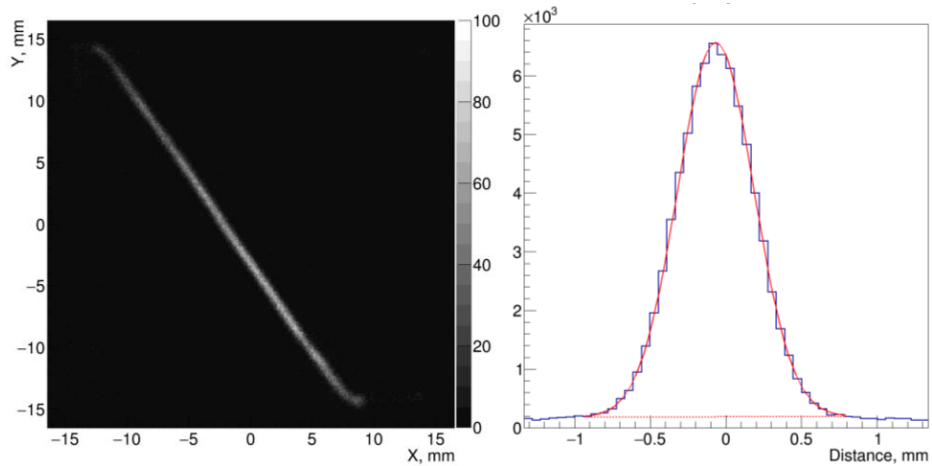

Figure 9. Left: Reconstructed image (ML method) for the dataset recorded with the cadmium mask with a 0.4 mm wide slit, rotated by 45 degrees. Right: The Gaussian fit of the image projection in the direction orthogonal to the slit gives 0.60 mm FWHM.

Flood and slit datasets were also recorded with a different detector configuration: the aluminium back reflector was replaced with a 1 mm thick PTFE plate, having somewhat higher and more diffuse light reflection (while resulting in more neutron scattering). The recorded SiPM signals were larger by 10%, however, the performance of the camera was essentially the same in terms of the spatial and energy resolution.

## 4   Conclusions and future work

In this study we experimentally demonstrate that a neutron Anger camera, based on a lithium-6 loaded scintillator and SiPM readout, can provide accurate reconstruction of scintillation events using a statistical method with LRFs obtained in an iterative reconstruction procedure. Characterization of the detector prototype at 2.5 Å neutron beam has demonstrated a useful field of view of 28 x 28 mm$^2$ for the glass scintillator and SiPM array size of 33.3 x 33.3 mm$^2$. The spatial resolution is better than 0.6 mm FWHM, and the energy resolution at the neutron peak is better than 11%.

The LRF reconstruction procedure can be performed in unsupervised mode giving the detector auto-calibration capabilities. Since only a flood field irradiation dataset is required for this calibration technique, application of the iterative method allows to avoid the standard calibration procedure based on a scan of the detector with a narrow pencil beam, which is impractical or even impossible to organize on a regular basis outside a specialized lab.

A dedicated study is planned on characterization of the gamma sensitivity with gamma ray sources. We also intend to replace GS20 with an organic or inorganic scintillator with a good capability for neutrons/gamma discrimination based on pulse shape analysis.

# 5  Acknowledgments

We would like to thank Dr. Bruno Guerard from ILL for his kind help in organization of the beam tests. This work was carried out with financial support from the Fundação para Ciência e Tecnologia (FCT) of Portugal through the project grants IF/00378/2013/CP1172/CT0001.